# Comparative Study of RF Heating in Deep Brain Stimulation Devices During MRI at 1.5 T and 0.55 T: Challenging the Assumption of Safety at Low Field Strengths.

Bhumi Bhusal[1], Pia Panravi Sanpitak[1,2], Jasmine Vu[1,2], Fuchang Jiang[1,2], Jacob Richardson[3], Nicole Seiberlich[3], Joshua M. Rosenow[4], Behzad Elahi[5], Laleh Golestanirad[1,2]

[1]*Department of Radiology, Northwestern University, Chicago, USA*
[2]*Department of Biomedical Engineering, Northwestern University, Evanston, USA*
[3]*Department of Radiology, University of Michigan, Ann Arbor, USA*
[4]*Department of Neurological Surgery, Northwestern University, USA*
[5]*Department of Neurology and Neurological Surgery, Loyola University Medical Center, USA*

Corresponding Author:
Laleh Golestanirad
737 N Michigan Ave, Suite 1600
Chicago, IL, 60611
Email: laleh.rad1@northwestern.edu

**Abstract:**

**Purpose**: Low-field MRI has been assumed to be implant-friendly based on limited studies. However, RF-induced heating due to an implant is a complex resonance phenomenon, highly dependent on the implant's configurations and the applied RF frequencies. This study aims to evaluate the RF heating of DBS implants during MRI at low-field strengths compared to higher field 1.5 T MRI.

**Methods**: A commercial deep brain stimulation (DBS) implant was used in full system as well as lead only configurations to evaluate and compare RF heating during MR imaging at 0.55 T and 1.5 T. The transfer function of the device at both configurations was measured and validated at each of the frequencies, which was then used for the in vivo prediction of RF heating for realistic DBS configurations at head, chest and abdomen imaging landmarks.

**Results**: For the lead only case, the RF heating due to the DBS was substantially smaller during imaging at 0.55 T compared to that at 1.5 T. However, for the full DBS system (longer implant), the RF heating at 0.55 T was comparable to and for some cases even higher than that at 1.5 T, reaching a level that poses risk of tissue damage in patients.

**Conclusions**: While RF heating generally tends to be lower at low-field MRI, the case with longer implanted leads demands extra caution, due to the higher possibility of matching resonant condition at low-field-strength frequencies. Thus, specific risk evaluation for each implant and configuration is required rather than assuming that lower field strength imaging is safer.

# Introduction

It is estimated that five year from now one-sixth of the global population will be 60 years of age or older, with a significant portion managing chronic diseases that require continuous monitoring and timely interventions [1]. This demographic shift is expected to drive an expansion of the market for active implantable medical devices (AIMDs) to a staggering $42 billion valuation by 2031 [2]. It is estimated that up to 75% of patients with AIMDs, such as deep brain stimulation (DBS) devices, will need at least one MRI exam during their lifetime, with many requiring multiple examinations [3]. This makes the issue of MRI accessibility for these individuals a pressing concern.

Over the past twenty years, substantial efforts have been made to make MRI accessible to patients with AIMDs, yet significant safety challenges persist. The major issue is the risk of radiofrequency (RF) heating of the tissue in the vicinity of implanted leads, a phenomenon commonly referred to as the "antenna effect," which has been proven significantly injurious in some cases [4]. To prevent injury, device manufacturers have established rigorous guidelines for MR conditionality. For instance, Abbott DBS devices are approved only at 1.5 T, with imaging limited to sequences with a $B_1^+ < 1.1$ µT for full DBS system with body transmit [5] and all systems with a capped extension wire (no attached DBS lead) are excluded from MRI. In many cases, these restrictions limit routine clinical sequences (see Table 3 in our recent work [6] for typical $B_1^+$ values of routine MRI exams), leading some groups to intentionally or unintentionally image patients off-label. There are reports of investigators imaging devices at unapproved strengths (mostly 3 T), during extended time windows, and with settings not verified in the MR guidelines [7, 8]. What is alarming about this now normalized practice is that safety assessments in these cases have been based on a handful of phantom experiments, falling far short of the recommended FDA testing procedures [9]. While the findings from such studies contribute to our understanding of treatments, they do not adequately address the risks of off-label use of MRI in patients with devices, creating a false sense of security for other investigators and clinicians, which could lead to unintended harm.

The recent advent of low-field MRI scanners (0.55 T and below) has introduced a fresh avenue for implant imaging. Their cost-effective nature and ease of installation make them ideal for deployment in remote, underserved regions, potentially leading to increased future adoption. With a lower field strength, susceptibility and off-resonance effects are reduced, thus the loss or degradation of signal is less severe around metallic implants [10-12]. Moreover, physical traction on ferromagnetic objects is greatly reduced at 0.55 T (by a factor of 7.5 compared to 1.5 T, and a factor of 30 compared to 3 T) making MRI safer in terms of device dislodgment. For these reasons, low-field scanners are publicized as implant-friendly [10, 13] despite the lack of robust data on their RF heating profile. This is alarming, because RF heating of an elongated implant is a resonance phenomenon, meaning that significant heating could occur when length of an implant approaches quarter or half-wavelength of the radiofrequency field in the tissue [14-16]. Therefore, theoretically, RF heating of a lead during low field MRI (e.g., 0.55 T) could be higher than at higher fields (e.g., 1.5 T and or 3 T), even for matched $B_1^+$ levels. If not taken into account, this phenomenon could cause serious thermal injuries in patients even at lower field strengths.

The goal of this study was to provide a rigorous analysis of RF heating of a commonly used deep brain stimulation (DBS) device (Infinity™ DBS system, Abbott Neuromodulation, Austin, TX) during MR imaging at 0.55 T and compare it to values observed during 1.5 T MRI. The DBS device was selected as it provides an excellent opportunity to assess two clinically relevant yet electromagnetically distinct RF heating scenarios. Specifically, DBS surgery is typically performed in two stages: first, leads (most commonly 40 cm in length) are implanted with their tips positioned within deep brain structures (e.g., basal ganglia) and the first 7-8 cm of the lead within the brain tissue. The extracranial portion of the lead exiting the surgical burr hole is positioned in the subgaleal space over the skull (see Supplementary Figure S1). Extra lead length is typically coiled posterior to the burr hole. Patients with a lead-only system would benefit from MRI for electrode localization and ruling out of complications [17]. In a subsequent surgery, the lead is connected to an extension cable of approximately 60 cm, which is routed subcutaneously in the posterior auricular region to an infraclavicular implantable pulse generator (IPG) in the chest (Supplementary Figure S1). Most MRI exams that are indicated for DBS patients are performed on those with a fully implanted system. The substantial difference in the length of the implanted conductive cable in a lead-only vs. fully implanted DBS system allows for the demonstration of the effect of MRI frequency and resonant length on RF heating.

In this work, we implemented the FDA-recommended approach outlined in Clause 8 of ISO-TS 10974 [9] to predict in vivo RF heating. Specifically, we measured and validated the transfer function of a commercial lead-only DBS system and a fully implanted system at both 1.5 T and 0.55 T, and used the validated transfer functions to predict in vivo RF heating of devices in a large cohort of patient models at both 1.5 T and 0.55 T. The concept of the device transfer function, originally introduced in 2007 [18] and refined over the past two decades [19-25] is currently mandated by the FDA for MR-conditional labeling [26] and provides a robust and unbiased prediction of in vivo RF heating. Additionally, to provide actionable information, we compiled look-up tables that present 95% confidence interval of temperature increase for each device configuration and imaging landmark, across a range of clinical $B_1^+$rms values. Our aim is to provide comprehensive data that can inform clinical decision-making and ensure patient safety. Additionally, our findings may offer insights for optimizing MRI protocols for patients with DBS devices, potentially expanding the accessibility and safety of MRI diagnostics for this patient population.

**Methods**

*Theory and Background*

Until 2007, the only method available to characterize RF-induced temperature increases around conductive implants during MRI was through in vitro testing, with implanted device placed in tissue-mimicking gel phantoms and exposed to MRI RF fields. However, this approach fell short of accurately predicting in vivo RF heating as the heterogeneity of human tissue properties substantially alters the distribution of MRI electric field [27, 28].

The advent of computational models that enabled the calculation of RF-induced electric fields in realistic heterogeneous body models [29, 30] spurred interest in mathematically modeling implants to evaluate in vivo RF heating. Given the complex internal structure of the lead, numerically solving the human and lead model concurrently is a daunting, if not insurmountable, task. A workaround solution, initially devised for large antenna systems, involves modelling the lead and the MRI-plus-human body separately and later merge the results to ascertain heating. In 2007, Park et al. elegantly formulated this technique specifically for MRI implant heating [18]. This concept, known as the lead's "transfer function", underpins the Tier 3 approach laid out in ISO-TS 10974, and has been meticulously refined over the past 15 years.

In brief, the in vivo RF-induced heating due to an AIMD can be calculated as:

$$\Delta T_{predicted} = C \, | \int_0^l TF(x) * E_{tan}(x) \, dx |^2$$

Where $TF(x)$ is the complex transfer function of the device (i.e., the validated mathematical model of the AIMD); $E_{tan}(x)$ is a complex number representing the magnitude and phase of the tangential component of the incident electric field in vivo along the trajectory of the lead, and the constant $C$ is a calibration constant which can be calculated using following expression:

$$C = \frac{\Delta T_{measured}}{\left| \int_0^l TF(x) * E_{tan}(x) \, dx \right|^2}$$

Here, $\Delta T_{measured}$ is the measured temperature increase for an arbitrary configuration of the device and $E_{tan}(x)$ is the tangential electric field extracted along the same configuration using electromagnetic modeling.

This framework elegantly estimates in vivo RF heating around implanted lead tips when the leads are routed along any arbitrary path in the human body: First, the lead's temperature response to a unit of locally applied electric field is measured on the bench when the lead is positioned straight in a homogeneous tissue-mimicking gel. This measurement constitutes the lead's uncalibrated transfer function. Next, the transfer function is calibrated and validated in a series of experiments, with the device configured along orthogonal trajectories that generate a wide range of RF heating [19, 31, 32]. Finaly, in vivo MRI incident electric fields are estimated via EM simulations with realistic heterogeneous body models but with no device in place. These simulations determine the electric field along any arbitrary trajectory passing through the human body when the subject undergoes MRI. Note that MRI-induced electric fields are a function of body position within the MRI coil and thus depend on the imaging landmark. Once the response of the lead to a unit of E field is known and the actual E field in the human body is determined, one can combine the two to predict how the lead would respond if it is routed along any arbitrary position within the human body.

*DBS device configurations*

We used a commercially available DBS device (Infinity™ DBS system, Abbott Laboratories, Abbott Park, IL) to evaluate and compare MRI-induced RF heating at a 0.55 T Siemens Free.Max scanner and a 1.5 T Siemens Aera scanner (Siemens Healthineers, Erlangen, Germany). The DBS device consisted of a 40 cm, 8-contact directional lead (model 6173), a 60 cm extension (model 6372), and an implantable pulse generator (IPG, model 6660). A fiber optic temperature sensor probe (OSENSA Inc; BC, Canada, resolution 0.01℃) was secured to the most distal electrode contact at the lead's tip. For the fully implanted case, the lead was connected to the extension and the IPG, and DBS device was set to MRI mode (stimulation off) as recommended by the manufacturer's guidelines. For the lead-only case, the lead was disconnected from the extension and capped at its proximal end using a manufacturer provided silicon cap.

*Transfer function measurement and validation*

We measured the transfer function of the DBS device using the reciprocal method described in Feng et al [33]. The measurement setup consisted of a two-port vector network analyzer (VNA 5063a, Keysight, Technologies USA), with one port connected to a current sensor and the other connected to a transmit probe (Figure 1A). The transmit probe was designed as a monopole antenna created from an RG402 semi-rigid coaxial cable with a 6 mm of its inner conductor exposed. The receive probe was also created using RG402 cable, with the inner conductor wound around a ferromagnetic torus (made of 4C65 ferrite) and soldered to its shield, similar to what was reported in previous studies [21, 24]. VNA ports were calibrated using the standard calibration kit (85032F, Keysight Technologies). The calibration plane was brought to the end of the coax-cables, where the cables connected to the receive and the transmit probes. The AIMD device was placed inside a rectangular box phantom filled with saline solution (conductivity σ =

0.5 S/m and relative permittivity $\varepsilon_r$ = 78) representing average tissue properties. The lead transfer function was then measured as the complex scattering parameter $S_{21}$ recorded while moving the receive probe along the length of the lead.

RF heating measurements for the purpose of transfer function calibration and validation were performed during MR imaging on a Siemens 0.55 T Free.Max scanner and a 1.5 T Aera scanner. Temperature increases at the DBS lead tip were recorded while the device was positioned inside an ASTM phantom filled with polyacrylic acid (PAA) gel (conductivity σ = 0.5 S/m and relative permittivity $\varepsilon_r$ = 86). The device was configured along orthogonal pathways that led to sufficient variation in the phase and magnitude of incident electric field [19] to generate a wide range of RF heating values. Device pathways utilized for the full DBS as well as lead-only systems are depicted in Figure 2. For both lead-only and full DBS systems, the transfer function was calibrated along an arbitrary trajectory which generated medium heating, while the remainder of the configurations were used for the validation of the model.

High SAR sequences with same acquisition time and $B_1^+$rms was used at both scanners. These included an axial T1-saturation pulse with acquisition time of 4:21 min, voxel size = 0.3 mm × 0.3 mm × 3.5 mm, TR = 671 ms, TE = 13 ms, FA = 180, and $B_1^+$rms = 4.5 µT at 0.55 T, and an axial T1-TSE sequence with acquisition time of 4:21 min, voxel size = 0.9 mm × 0.9 mm × 1.4 mm, TR = 836 ms, TE = 7.3 ms, FA = 165, and $B_1^+$rms = 4.5 µT at 1.5 T. The phantom was scanned in a head-first supine position in both scanners, with the coil iso-center coinciding a transverse plane 5 cm inferior to the shoulder.

***Numerical simulations with realistic body model***

Numerical simulations were implemented using HFSS in Ansys Electronic Desktop 2021 (Ansys Inc., Canonsburg, PA, USA). Models of birdcage body transmit coils were created to represent Siemens 1.5 T Aera and 0.55 T Free.Max body transmit coils based on the information provided by the manufacturer. The 1.5 T coil was modeled as a 16 rung high-pass coil, tuned at 63.6 MHz and driven in quadrature mode by two sinusoidal sources with 90° separation in phase and position along the back end-ring. The 0.55 T coil was modeled as a 16 rung low-pass coil, tuned at 23.6 MHz and driven in quadrature mode with sources in similar position.

The Ansys Human Body Model (HBM), consisting of over 300 body parts, was used for numerical modeling [34]. The model represented an adult male with a height of 180 cm and was assigned frequency-dependent tissue dielectric properties. Table S1 in the Supplementary Material provides the dielectric properties of different tissue classes at 63.6 MHz and 23.6 MHz.

The trajectories of full DBS devices as well as lead only cases were created taking reference of the post-operative CT images of the patients with DBS implants. The use of post-operative CT images of the patient was approved by the institutional review board (IRB). The number of trajectories was augmented by manually adding extra variation in the size and position of the loops, as well as the trajectory path along the skull using a computer-aided design (CAD) tool (Rhino 7.0, Robert McNeal and Associates, Seattle, WA). This represented additional possible configurations of the device, which can vary based on the surgeon's preference and practices. In total, we created 210 trajectories of full DBS devices and 210 trajectories of lead-only cases. The configurations of lead-only cases were created from the full DBS configurations by removing the extension cable, thus shortening the trajectories to a length of 42.5 cm.

The body model, along with the ensemble of lead trajectories, was then positioned inside each of the body transmit coils in a head-first supine position and solved for the electromagnetic fields. The ANSYS HFSS followed an adaptive mesh setting, where it performed successive refinement of an initial mesh between the iterative passes. The convergence criteria for each of the simulations were set in terms of the maximum change in the scattering parameters (ΔS), defined as $Max_{ij}|S_{ij}^N - S_{ij}^{N-1}|$, where $i$ and $j$ cover all ports and N represents the number of iterative passes. The simulations were considered to have converged when the maximum value of ΔS ≤ 0.02 was obtained. Simulations were repeated for each case at three different imaging landmarks, with the head, chest, and abdomen positioned at the coil's iso-center. The details of the body model, coil models, device trajectory configurations, as well as the landmarks, are depicted in Figure 3.

For each trajectory configuration, the complex value of the tangential component of the E-field along the trajectory path was extracted for the body positioned at each of the three imaging landmarks. The input power of coils was adjusted so that the mean value of $B_1^+$ on a 5-cm diameter plane passing through the coil iso-center was 4.5 μT. The RF heating was then predicted using a calibrated transfer function and the complex value of the tangential E-field for each of the device configurations.

**Results**

*DBS device transfer function and phantom experiments*

The magnitude and phase of the transfer functions of full DBS device and lead only system at 64 MHz (1.5 T) and 23.6 MHz (0.55 T) are depicted in Figure 4. Figure 5 shows the results of the TF-predicted temperature rise compared to measured values. The 95% confidence intervals (CI) of the difference in predicted and measured temperature increase ($\Delta T_{predicted} - \Delta T_{measured}$) were [0.21°C-7.96°C] and [0.05°C-0.55°C] for the lead only case at 1.5 T and 0.55 T respectively, and were [0.67°C-3.99°C] and [0.87°C-4.90°C] for the full system at 1.5 T and 0.55 T, respectively. Since the gel temperature cannot rise beyond 100°C (the boiling point of water), the upper limit for predicted temperature was set at 100°C.

Although the TF approach allows for statistical inferences on in vivo RF heating in anatomically realistic populations, it is still insightful to compare the measured RF heating for identical device configurations under the same level of RF exposure in a simple phantom. We plotted the temperature increase measured for the configurations shown in Figure 2, as depicted in the bar plots in Figure 6. It is clear from the plots that the temperature increase at the 0.55 T scanner is consistently and substantially smaller than that at 1.5 T for the DBS lead-only case. However, for the full DBS system, the RF heating at 0.55 T is comparable to, and in some cases even surpasses, the heating at 1.5 T, highlighting the importance of considering the conductive component's length in RF heating.

Figure 7 shows the absolute magnitude of the tangential component of the incident electric field along the length of the lead and extension of the full DBS device when routed along trajectory ID 1 (see Figure 2). The input power of both coils was adjusted to produce same value of $B_1^+$rms (4.5 µT) on an axial plane passing through the coil iso-center. Notably, the magnitude of the incident E-field at 0.55 T is reduced by approximately a factor of 3 compared to that at 1.5 T, as it is expected to be proportional to the reduction in frequency (23.6 MHz vs 63.6 MHz). However, the RF heating was interestingly slightly higher at 0.55 T than at 1.5 T for the same configuration (see Figure 6), highlighting the effect of length-dependent resonance.

***In vivo prediction of RF heating***

The in vivo RF heating predicted by the transfer function is presented in violin plots of Figure 8 for the 210 full DBS system and the 210 lead only cases at both 1.5 T and 0.55 T and at three different imaging landmarks corresponding to head, chest and abdomen.

For the full DBS system, RF heating was highest at the head and chest imaging landmarks and lowest at the abdomen imaging landmark across both scanners. At 0.55 T, the mean ± standard deviation of RF heating was 3.22 °C ± 1.64 °C at the head, 2.15 °C ± 1.52 °C at the chest, and 0.35 °C ± 0.09 °C at the abdomen. At 1.5 T, the corresponding values were 5.02 °C ± 4.78 °C at the head, 3.94 °C ± 4.04 °C at the chest, and 0.62 °C ± 0.20 °C at the abdomen.

For the lead-only case, RF heating was highest at the head imaging landmark and significantly lower at the chest and abdomen imaging landmarks across both scanners. The mean ± standard deviation of RF heating for the lead-only cases was 0.21 °C ± 0.17 °C at 0.55 T and 3.30 °C ± 2.47 °C at 1.5 T for the head imaging landmark, indicating more than a 15-fold reduction in average heating at 0.55 T. For the chest and abdomen landmarks, both the mean and standard deviation of RF heating were below 0.2 °C. Unlike the full DBS system, where the average RF heating at 0.55 T was comparable to that at 1.5 T, the lead-only system exhibited negligible heating at 0.55 T compared to 1.5 T.

We used the Shapiro-Wilk normality test to analyze the distribution of RF heating at head, chest, and abdomen landmarks for each of the field strengths. In the case of the full DBS system, the temperature rise distribution showed a non-normal distribution for all landmarks at each field strength ($p < 0.001$). Additionally, a Wilcoxon signed-rank test with a significance level of $\alpha=0.01$ indicated a significantly lower temperature rise at 0.55 T compared to that at 1.5 T for all three imaging landmarks ($p < 10^{-6}$ for the head and chest landmarks, and $p < 10^{-25}$ for the abdomen landmark). For DBS lead-only case, the distribution of temperature increase at the head imaging landmark also showed a non-normal distribution ($p < 10^{-4}$). The Wilcoxon signed-rank test indicated significantly lower heating at 0.55 T compared to 1.5 T ($p < 10^{-30}$). The temperature increases for lead-only cases at the chest and abdomen landmarks were negligible and were not statistically analyzed.

Note that the results correspond to the maximum temperature increase after 4:20 minutes of continuous scanning at a $B_1^+$rms of 4.5 µT, which is close to the maximum allowed $B_1^+$rms for both scanners. In most cases (with temperature increases of less than 15°C, close to the upper limit of observed in vivo predicted temperature increase in this study) the temperature rise plateaus after 5 minutes of scanning, meaning the temperature increases by less than 0.1°C every 30 seconds after the 5 minutes. Therefore, the scan time was considered reasonable to provide a conservative estimation of the risk in realistic scenarios. For sequences with a lower $B_1^+$rms, scaling according to the square of the $B_1^+$rms ratios would yield a close estimate of RF heating. That is the ratio of temperature increase, $\Delta T_2/\Delta T_1 = (B_2/B_1)^2$. Figure S2 in the Supplementary Material shows the violin plots of predicted in vivo temperature increases for $B_1^+$rms values in the range of 0.5 µT to 4 µT. Table S2 in the Supplementary Material presents the 95% confidence interval of temperature increase for each device configuration and imaging landmark, across the same range of $B_1^+$rms values. The confidence interval was calculated as the interval represented by 2.5[th] percentile and the 97.5[th] percentile value of the temperature increase for each case. This method represented the most conservative estimate for the upper limit of the temperature increase and was chosen to provide an additional safety margin to ensure the patient's welfare.

**Discussion**

Over the past decade, there has been a significant increase in efforts to make MRI accessible to patients with active implantable medical devices (AIMDs). Several research groups have focused on field shaping methods, where the electric field of MRI transmit coils is tailored to create a steerable region of low electric field that aligns with the patient's implant [35-43]. Another promising approach involves modifying the implant trajectory itself to minimize its interaction with the MRI electric fields [21, 44-46]. Additionally, new implant materials and geometries have been explored [47-50]. While these techniques show promise, they have not yet been widely adopted in clinical practice. There have also been preliminary, promising reports from scanners with alternative field polarizations, such as vertical open bore scanners [42, 51, 52], although a robust analysis across large population models is still lacking.

The excitement surrounding the new generation of low-field MRI scanners is driven in part by their potential for greater compatibility with implants. While this is true in terms of reduced metal artifacts and attractive forces, there is limited data available on the RF safety profile of these scanners.

In this study, we examined the RF heating of a commercial DBS device under various clinically relevant conditions. AIMD models were developed and validated based on measurements of the device transfer function at frequencies relevant to 0.55 T and 1.5 T MRI. These models were then used to predict RF heating in realistic human body configurations, across a range of imaging landmarks and $B_1^+$rms values.

As expected, RF-induced heating was significantly dependent on the length of the implant. Overall, predicted RF heating was lower at 0.55 T compared to 1.5 T for both lead-only and full DBS systems across all imaging landmarks. However, caution was warranted based on implant lengths. Specifically, the full DBS system with a longer conductive cable (lead + extension) exhibited significant RF heating at 0.55 T, potentially reaching damaging levels and surpassing heating at 1.5 T in certain configurations. Conversely, RF heating for the lead-only system at 0.55 T was well below the risk level and substantially lower than at 1.5 T. These findings underscored the need for caution when scanning patients with elongated medical implants using low-field scanners, as generalizations about lower risk may not always be accurate.

A limited number of prior studies suggested that low-field MRI scanning to be relatively safer for certain interventional catheters [53, 54]. However, results from our study emphasize that RF heating at low-field scanners can be significantly lower compared to 1.5 T for short implants (such as DBS lead-only systems), but comparable to or higher than that at 1.5 T for longer implants (such as DBS full systems). This finding aligns with previous simulation studies showing that resonant heating of leads at low fields can exceed that at 1.5 T under certain conditions [55]. This underscores the need for individual risk assessments rather than blanket generalizations.

***The Myth of Safety at Low Fields***

In the realm of MRI safety, a crucial metric often serving as a proxy for RF heating is the Specific Absorption Rate (SAR), which quantifies the amount of power deposited in a given mass of tissue by the RF excitation. Accurately calculating local SAR in human subjects is a complex task, demanding sophisticated mathematical modeling and numerical simulations. However, a valuable equation can be derived from a simplified model, as follows [56]:

$$SAR = (\sigma A^2 \omega^2 \boldsymbol{B}_1^2 D)/2\rho. \quad \text{Eq (1)}$$

Here σ represents tissue conductivity, A stands for the cross-section of the body, ω denotes the RF frequency, D accounts for the pulse duty cycle, and ρ corresponds to tissue density. Equation (1) suggests that, all else being equal, SAR tends to be lower at lower magnetic field strengths due to a lower ω, which has been also demonstrated here (see Figure 7). This observation has fueled a common misconception that low-field MRI scanners are generally safer in terms of RF heating—an assertion that holds true in the absence of metallic implants. However, it's crucial to recognize that the presence of an elongated implant, such as a lead in neuromodulation or cardiovascular devices, can amplify SAR at its tip and lead to increased heating.

The extent of SAR amplification hinges on the ratio of the lead's length to the RF field wavelength within the tissue, position of implant inside body, as well as the dielectric properties of the tissue/media surrounding the implant [6, 41, 57-63]. Previous studies have shown several-fold variation in RF heating due to an implant under constant RF exposure when its length was varied [14-16]. These studies suggest that when the length of the implant's lead matches certain multiple of the wavelength of the incident RF field, the heating at the lead tip can be dramatically amplified compared to other lengths.

Low-field scanners operate at lower RF frequencies, resulting in longer wavelength of the RF fields. As a result, longer implants are more likely to produce resonant RF heating at these field strengths, leading to amplification of RF heating. This phenomenon is clearly evidenced in our study: the predicted RF heating of the DBS lead only (shorter implant) was substantially lower at 0.55 T than at 1.5 T, whereas the RF heating of the full DBS (longer implant) was comparable to— and for certain cases even higher than—that at 1.5 T.

The predicted results indicate that in certain instances, RF heating at 0.55 T exceeded critically high levels (> 6 °C). A temperature increase beyond 6°C is capable of producing a thermal dose that surpasses the threshold value for brain tissue damage (~10 CEM 43°C) within first 10 minutes of continuous RF exposure [64]. This implies that the RF heating safety of implants on low-field scanners should be assessed on a case-by-case basis depending upon the implant's configuration. The findings emphasized that blanket safety assertions about low-field MRI may be misleading, necessitating careful consideration based on specific clinical contexts, medical devices, and patient factors similar to statements made in a previous article [65].

*The Devil is in the Generalization*

Although as mentioned above, our collective knowledge drawn from numerous studies over the past twenty years supports the conclusion that shorter leads should generate less heating in low-field scanners compared to higher fields, the concept of "length" can be misleading, as the helical winding of internal wires in most AIMD leads means the true electrical length of the implant may be substantially longer than its apparent physical length. This is illustrated in Figure 9, which shows the mammogram of two 40-cm DBS leads from different manufacturers. As shown, the pitch of the internal microwire connecting the electrode contacts to the IPG varies significantly between the two leads. This indicates that seemingly similar leads can have vastly different electrical lengths, and thus, results from one lead model cannot be generalized to another.

## Acknowledgements

This work was supported by National Institute of Health (NIH) grants R01EB036272 and R03EB033864. Additionally, the authors would like the acknowledge the in-kind donation of DBS devices from Abbott, and the Center for Translational Imaging (CTI) at Northwestern University for the availability of MRI scanners for this study.

## References

[1]     WHO. "Aging and Health (https://www.who.int/news-room/fact-sheets/detail/ageing-and-health#:~:text=By%202030%2C%201%20in%206,will%20double%20(2.1%20billion)." (accessed.
[2]     "Active Implantable Medical Devices Market Size, Share & Trends Analysis Report Frecasts, 2023-2031." https://straitsresearch.com/report/active-implantable-medical-devices-market#:~:text=Market%20Overview,USD%2022.65%20billion%20in%202022 (accessed.

[3]     S. Falowski, Y. Safriel, M. P. Ryan, and L. Hargens, "The Rate of Magnetic Resonance Imaging in Patients with Deep Brain Stimulation," *Stereotact Funct Neurosurg,* vol. 94, no. 3, pp. 147-53, 2016, doi: 10.1159/000444760.

[4]     J. M. Henderson, J. Tkach, M. Phillips, K. Baker, F. G. Shellock, and A. R. Rezai, "Permanent neurological deficit related to magnetic resonance imaging in a patient with implanted deep brain stimulation electrodes for Parkinson's disease: case report," *Neurosurgery,* vol. 57, no. 5, p. E1063, 2005.

[5]     Abbott. (2024) MRI Procedure Information. Available: https://manuals.eifu.abbott/en/index.html

[6]     B. T. Nguyen *et al.*, "Safety of MRI in patients with retained cardiac leads," *Magn Reson Med,* vol. 87, no. 5, pp. 2464-2480, May 2022, doi: 10.1002/mrm.29116.

[7]     A. Boutet *et al.*, "3-Tesla MRI of deep brain stimulation patients: safety assessment of coils and pulse sequences," *J Neurosurg,* vol. 132, no. 2, pp. 586-594, Feb 22 2019, doi: 10.3171/2018.11.JNS181338.

[8]     B. Li, A. Boutet, and A. M. Lozano, "Scanning Contraindicated Deep Brain Stimulator Patients on 3 Tesla MRI–A Single Centre Experience," *Journal of Medical Imaging and Radiation Sciences,* vol. 50, no. 3, p. S4, 2019.

[9]     ISO/TS 10974.  International Organization of Standarization, "Assessment of the safety of magnetic resonance imaging for patients with an active implantable medical device," 2018.

[10]    A. E. Campbell-Washburn *et al.*, "Opportunities in interventional and diagnostic imaging by using high-performance low-field-strength MRI," *Radiology,* vol. 293, no. 2, pp. 384-393, 2019.

[11]    R. Heiss, A. M. Nagel, F. B. Laun, M. Uder, and S. Bickelhaupt, "Low-field magnetic resonance imaging: a new generation of breakthrough technology in clinical imaging," *Investigative radiology,* vol. 56, no. 11, pp. 726-733, 2021.

[12]    I. Khodarahmi *et al.*, "New-generation low-field magnetic resonance imaging of hip arthroplasty implants using slice encoding for metal artifact correction: first in vitro experience at 0.55 T and comparison with 1.5 T," *Investigative radiology,* vol. 57, no. 8, pp. 517-526, 2022.

[13]    B. Ahmed Khan and E. L. Siegel, "Could Very Low Field Strength Be the Next Frontier for MRI?," *Diagnostic Imaging* 2021.

[14]    C. J. Yeung, P. Karmarkar, and E. R. McVeigh, "Minimizing RF heating of conducting wires in MRI," *Magn Reson Med,* vol. 58, no. 5, pp. 1028-34, Nov 2007, doi: 10.1002/mrm.21410.

[15]    B. Bhusal, P. Bhattacharyya, T. Baig, S. Jones, and M. Martens, "Measurements and simulation of RF heating of implanted stereo-electroencephalography electrodes during MR scans," *Magn Reson Med,* vol. 80, no. 4, pp. 1676-1685, Oct 2018, doi: 10.1002/mrm.27144.

[16]    C. Armenean, E. Perrin, M. Armenean, O. Beuf, F. Pilleul, and H. Saint-Jalmes, "RF-induced temperature elevation along metallic wires in clinical magnetic resonance imaging: influence of diameter and length," *Magn Reson Med,* vol. 52, no. 5, pp. 1200-6, Nov 2004, doi: 10.1002/mrm.20246.

[17]    A. Horn, "The impact of modern-day neuroimaging on the field of deep brain stimulation," *Curr Opin Neurol,* vol. 32, no. 4, pp. 511-520, Aug 2019, doi: 10.1097/WCO.0000000000000679.

[18]    S. M. Park, R. Kamondetdacha, and J. A. Nyenhuis, "Calculation of MRI-induced heating of an implanted medical lead wire with an electric field transfer function," *J Magn Reson Imaging,* vol. 26, no. 5, pp. 1278-85, Nov 2007, doi: 10.1002/jmri.21159.

[19]    J. F. Zheng *et al.*, "Developing AIMD Models Using Orthogonal Pathways for MRI Safety Assessment," (in English), *Ieee Transactions on Electromagnetic Compatibility,* vol. 62, no. 6, pp. 2689-2695, Dec 2020, doi: 10.1109/Temc.2020.2997236.

[20] S. Feng, R. Qiang, W. Kainz, and J. Chen, "A technique to evaluate MRI-induced electric fields at the ends of practical implanted lead," *IEEE Transactions on microwave theory and techniques,* vol. 63, no. 1, pp. 305-313, 2014.

[21] F. Jiang et al., "Modifying the trajectory of epicardial leads can substantially reduce MRI-induced RF heating in pediatric patients with a cardiac implantable electronic device at 1.5 T," *Magnetic resonance in medicine,* vol. 90, no. 6, pp. 2510-2523, 2023.

[22] E. Mattei, F. Censi, G. Calcagnini, E. Lucano, and L. M. Angelone, "A combined computational and experimental approach to assess the transfer function of real pacemaker leads for MR radiofrequency-induced heating," *MAGMA,* vol. 34, no. 4, pp. 619-630, Aug 2021, doi: 10.1007/s10334-021-00909-0.

[23] J. Kabil, J. Felblinger, P. A. Vuissoz, and A. Missoffe, "Coupled transfer function model for the evaluation of implanted cables safety in MRI," *Magn Reson Med,* vol. 84, no. 2, pp. 991-999, Aug 2020, doi: 10.1002/mrm.28146.

[24] A. Missoffe and S. Aissani, "Experimental setup for transfer function measurement to assess RF heating of medical leads in MRI: Validation in the case of a single wire," *Magn Reson Med,* vol. 79, no. 3, pp. 1766-1772, Mar 2018, doi: 10.1002/mrm.26773.

[25] T. Lottner, S. Reiss, A. Bitzer, M. Bock, and A. C. Özen, "A transfer function measurement setup with an electro-optic sensor for MR safety assessment in cascaded media," *IEEE Transactions on Electromagnetic Compatibility,* vol. 63, no. 3, pp. 662-672, 2020.

[26] "US Food and Drug Administration." https://www.fda.gov/regulatory-information/search-fda-guidance-documents/testing-and-labeling-medical-devices-safety-magnetic-resonance-mr-environment (accessed.

[27] C. M. Collins, S. Li, and M. B. Smith, "SAR and B1 field distributions in a heterogeneous human head model within a birdcage coil. Specific energy absorption rate," (in eng), *Magn Reson Med,* vol. 40, no. 6, pp. 847-56, Dec 1998, doi: 10.1002/mrm.1910400610.

[28] E. Gjonaj, M. Bartsch, M. Clemens, S. Schupp, and T. Weiland, "High-resolution human anatomy models for advanced electromagnetic field computations," *IEEE Transactions on Magnetics,* vol. 38, no. 2, pp. 357-360, 2002, doi: 10.1109/20.996096.

[29] C. M. Collins et al., "Temperature and SAR calculations for a human head within volume and surface coils at 64 and 300 MHz," (in eng), *J Magn Reson Imaging,* vol. 19, no. 5, pp. 650-6, May 2004, doi: 10.1002/jmri.20041.

[30] H. S. Ho, "Safety of metallic implants in magnetic resonance imaging," (in eng), *J Magn Reson Imaging,* vol. 14, no. 4, pp. 472-7, Oct 2001, doi: 10.1002/jmri.1209.

[31] Q. Zeng, Q. Wang, J. Zheng, W. Kainz, and J. Chen, "Evaluation of MRI RF electromagnetic field induced heating near leads of cochlear implants," *Phys Med Biol,* vol. 63, no. 13, p. 135020, Jul 6 2018, doi: 10.1088/1361-6560/aacbf2.

[32] Q. Zeng et al., "Investigation of RF-induced heating near interventional catheters at 1.5 T MRI: A combined modeling and experimental study," *IEEE Transactions on Electromagnetic Compatibility,* vol. 61, no. 5, pp. 1423-1431, 2018.

[33] S. Feng, R. Qiang, W. Kainz, and J. Chen, "A technique to evaluate MRI-induced electric fields at the ends of practical implanted lead," *IEEE Transactions on microwave theory techniques,* vol. 63, no. 1, pp. 305-313, 2014.

[34] "ANSYS Human Body Model." https://catalog.ansys.com/product/6414aa87bb452e0b3f107620/ansys-human-body-m (accessed.

[35] B. Guerin, L. M. Angelone, D. Dougherty, and L. L. Wald, "Parallel transmission to reduce absorbed power around deep brain stimulation devices in MRI: Impact of number and

[35] arrangement of transmit channels," *Magn Reson Med,* vol. 83, no. 1, pp. 299-311, Jan 2020, doi: 10.1002/mrm.27905.

[36] Y. Eryaman, E. A. Turk, C. Oto, O. Algin, and E. Atalar, "Reduction of the radiofrequency heating of metallic devices using a dual-drive birdcage coil," *Magn Reson Med,* vol. 69, no. 3, pp. 845-52, Mar 1 2013, doi: 10.1002/mrm.24316.

[37] C. E. McElcheran, B. S. Yang, K. J. Anderson, L. Golestanirad, and S. J. Graham, "Parallel radiofrequency transmission at 3 tesla to improve safety in bilateral implanted wires in a heterogeneous model," (in English), *Magnetic Resonance in Medicine,* vol. 78, no. 6, pp. 2406-2415, Dec 2017, doi: 10.1002/mrm.26622.

[38] C. McElcheran *et al.*, "Parallel transmission for heating reduction in realistic deep brain stimulation lead trajectories," in *Proc. Intl. Soc. Mag. Reson. Med*, 2017, vol. 25.

[39] L. Golestanirad *et al.*, "Construction and modeling of a reconfigurable MRI coil for lowering SAR in patients with deep brain stimulation implants," *Neuroimage,* vol. 147, pp. 577-588, Feb 15 2017, doi: 10.1016/j.neuroimage.2016.12.056.

[40] L. Golestanirad *et al.*, "Reconfigurable MRI coil technology can substantially reduce RF heating of deep brain stimulation implants: First in-vitro study of RF heating reduction in bilateral DBS leads at 1.5 T," (in English), *Plos One,* vol. 14, no. 8, p. e0220043, Aug 7 2019, doi: ARTN e022004310.1371/journal.pone.0220043.

[41] E. Kazemivalipour *et al.*, "Reconfigurable MRI technology for low-SAR imaging of deep brain stimulation at 3T: Application in bilateral leads, fully-implanted systems, and surgically modified lead trajectories," (in English), *Neuroimage,* vol. 199, pp. 18-29, Oct 1 2019, doi: 10.1016/j.neuroimage.2019.05.015.

[42] E. Kazemivalipour *et al.*, "Vertical open-bore MRI scanners generate significantly less radiofrequency heating around implanted leads: A study of deep brain stimulation implants in 1.2 T OASIS scanners versus 1.5 T horizontal systems," *Magnetic Resonance in Medicine,* vol. 86, no. 3, pp. 1560-1572, 2021.

[43] M. Etezadi-Amoli, P. Stang, A. Kerr, J. Pauly, and G. Scott, "Controlling radiofrequency-induced currents in guidewires using parallel transmit," *Magnetic resonance in medicine,* vol. 74, no. 6, pp. 1790-1802, 2015.

[44] J. Vu, B. Bhusal, J. M. Rosenow, J. Pilitsis, and L. Golestanirad, "Effect of surgical modification of deep brain stimulation lead trajectories on radiofrequency heating during MRI at 3T: from phantom experiments to clinical implementation," *Journal of neurosurgery,* vol. 140, no. 5, pp. 1459-1470, 2023.

[45] B. Bhusal *et al.*, "Effect of Device Configuration and Patient's Body Composition on the RF Heating and Nonsusceptibility Artifact of Deep Brain Stimulation Implants During MRI at 1.5 T and 3T," vol. 53, p. 11, 2020.

[46] L. Golestanirad, L. M. Angelone, M. I. Iacono, H. Katnani, L. L. Wald, and G. Bonmassar, "Local SAR near deep brain stimulation (DBS) electrodes at 64 and 127 MH z: A simulation study of the effect of extracranial loops," *Magnetic resonance in medicine,* vol. 78, no. 4, pp. 1558-1565, 2017.

[47] L. Golestanirad *et al.*, "Reducing RF-Induced Heating Near Implanted Leads Through High-Dielectric Capacitive Bleeding of Current (CBLOC)," (in English), *Ieee Transactions on Microwave Theory and Techniques,* vol. 67, no. 3, pp. 1265-1273, Mar 2019, doi: 10.1109/Tmtt.2018.2885517.

[48] T. Zaidi, F. Marturano, G. Bonmassar, and L. Golestanirad, "Reduction of Medical Device Heating During Mri At 1.5 T and 3 T: Design and Experimental Validation of A New Lead Construct," in *2024 IEEE International Symposium on Biomedical Imaging (ISBI)*, 2024: IEEE, pp. 1-5.

[49] R. W. Gray, W. T. Bibens, and F. G. Shellock, "Simple design changes to wires to substantially reduce MRI-induced heating at 1.5 T: implications for implanted leads," *Magn Reson Imaging,* vol. 23, no. 8, pp. 887-891, 2005.

[50] Y. Wang *et al.*, "A technique for the reduction of RF-induced heating of active implantable medical devices during MRI," *Magn Reson Med,* vol. 87, no. 1, pp. 349-364, Jan 2022, doi: 10.1002/mrm.28953.

[51] L. Golestanirad *et al.*, "RF heating of deep brain stimulation implants in open-bore vertical MRI systems: A simulation study with realistic device configurations," *Magn Reson Med,* vol. 83, no. 6, pp. 2284-2292, Jun 2020, doi: 10.1002/mrm.28049.

[52] J. Vu *et al.*, "A comparative study of RF heating of deep brain stimulation devices in vertical vs. horizontal MRI systems," *Plos one,* vol. 17, no. 12, p. e0278187, 2022, doi: 10.1371/journal.pone.0278187.

[53] A. C. Özen *et al.*, "RF-induced heating of interventional devices at 23.66 MHz," *Magnetic Resonance Materials in Physics, Biology Medicine,* pp. 1-11, 2023.

[54] A. E. Campbell-Washburn *et al.*, "Opportunities in Interventional and Diagnostic Imaging by Using High-Performance Low-Field-Strength MRI," *Radiology,* vol. 293, no. 2, pp. 384-393, Nov 2019, doi: 10.1148/radiol.2019190452.

[55] P. Sanpitak, B. Bhusal, J. Vu, and L. Golestanirad, "Low-field MRI's Spark on Implant Safety: A Closer Look at Radiofrequency Heating," in *2023 45th Annual International Conference of the IEEE Engineering in Medicine & Biology Society (EMBC)*, 2023: IEEE, pp. 1-5.

[56] M. Tang and T. Yamamoto, "Progress in understanding radiofrequency heating and burn injuries for safer MR imaging," *Magnetic Resonance in Medical Sciences,* vol. 22, no. 1, pp. 7-25, 2023.

[57] B. Bhusal, B. Keil, J. Rosenow, E. Kazemivalipour, and L. Golestanirad, "Patient's body composition can significantly affect RF power deposition in the tissue around DBS implants: ramifications for lead management strategies and MRI field-shaping techniques," (in English), *Physics in Medicine and Biology,* vol. 66, no. 1, p. 015008, Jan 7 2021, doi: ARTN 01500810.1088/1361-6560/abcde9.

[58] B. Bhusal *et al.*, "Safety and image quality at 7T MRI for deep brain stimulation systems: Ex vivo study with lead-only and full-systems," (in English), *Plos One,* vol. 16, no. 9, p. e0257077, Sep 7 2021, doi: ARTN e025707710.1371/journal.pone.0257077.

[59] F. Jiang *et al.*, "A comparative study of MRI-induced RF heating in pediatric and adult populations with epicardial and endocardial implantable electronic devices," in *2022 44th Annual International Conference of the IEEE Engineering in Medicine & Biology Society (EMBC)*, 2022: IEEE, pp. 4014-4017.

[60] C. E. McElcheran *et al.*, "Numerical Simulations of Realistic Lead Trajectories and an Experimental Verification Support the Efficacy of Parallel Radiofrequency Transmission to Reduce Heating of Deep Brain Stimulation Implants during MRI," *Sci Rep,* vol. 9, no. 1, p. 2124, Feb 14 2019, doi: 10.1038/s41598-018-38099-w.

[61] J. A. Martinez, P. Serano, and D. B. Ennis, "Patient Orientation Affects Lead-Tip Heating of Cardiac Active Implantable Medical Devices during MRI," *Radiol Cardiothorac Imaging,* vol. 1, no. 3, p. e190006, Aug 29 2019, doi: 10.1148/ryct.2019190006.

[62] L. Alon, C. M. Deniz, G. Carluccio, R. Brown, D. K. Sodickson, and C. M. Collins, "Effects of Anatomical Differences on Electromagnetic Fields, SAR, and Temperature Change," *Concepts Magn Reson Part B Magn Reson Eng,* vol. 46, no. 1, pp. 8-18, Feb 1 2016, doi: 10.1002/cmr.b.21317.

[63]  B. Bhusal, P. Bhattacharyya, T. Baig, S. Jones, and M. Martens, "Effect of inter-electrode RF coupling on heating patterns of wire-like conducting implants in MRI," *Magn Reson Med,* vol. 87, no. 6, pp. 2933-2946, Jun 2022, doi: 10.1002/mrm.29177.
[64]  G. C. Van Rhoon, T. Samaras, P. S. Yarmolenko, M. W. Dewhirst, E. Neufeld, and N. Kuster, "CEM43° C thermal dose thresholds: a potential guide for magnetic resonance radiofrequency exposure levels?," *European radiology,* vol. 23, pp. 2215-2227, 2013.
[65]  T. Gilk and E. Kanal, "MRI safety considerations associated with low-field MRI: mostly good news," (in English), *Magnetic Resonance Materials in Physics Biology and Medicine,* vol. 36, no. 3, pp. 427-428, Jul 2023, doi: 10.1007/s10334-023-01079-x.

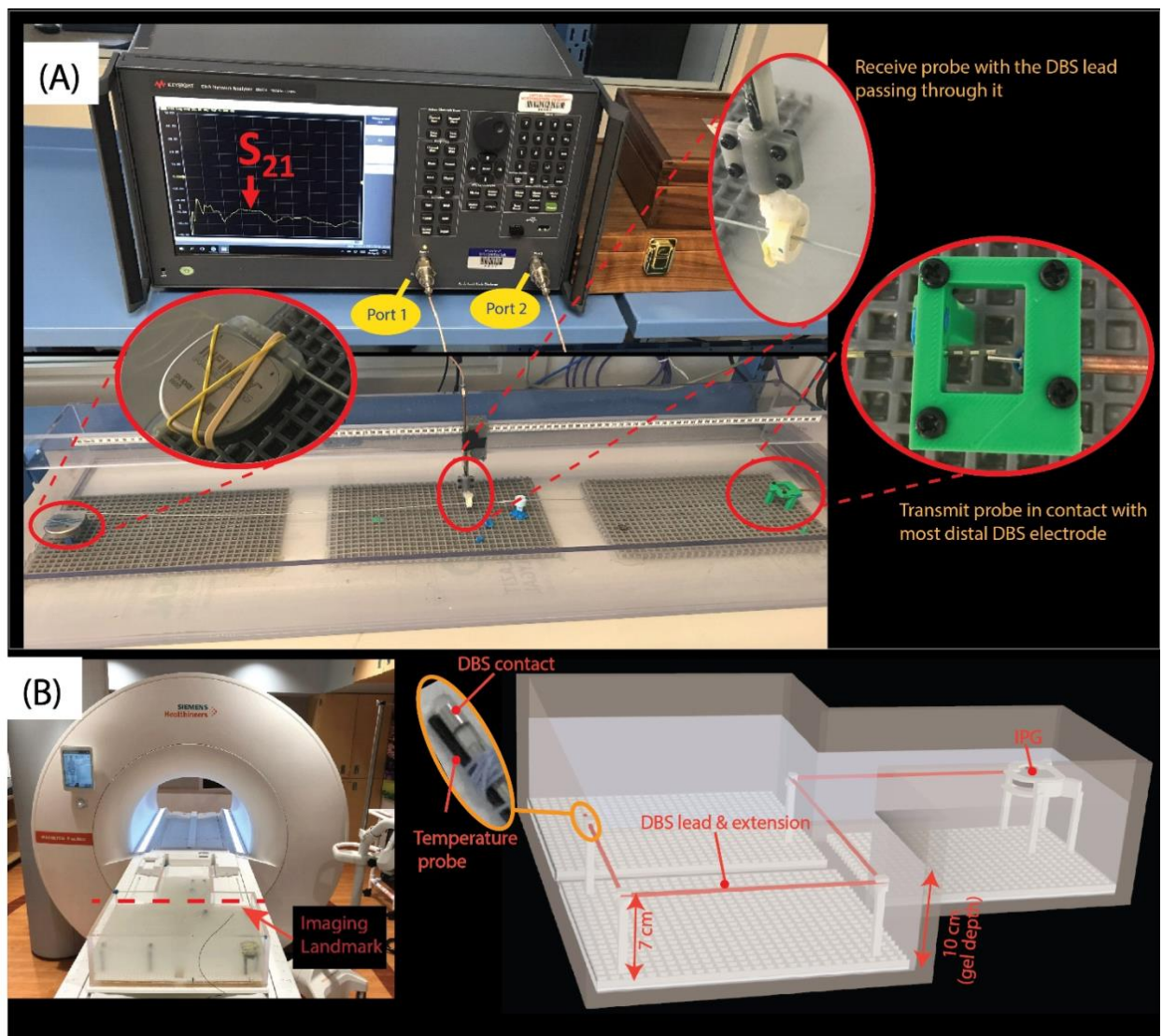

Figure 1: (A)Transfer function measurement setup with close view of the receive probe, DBS device lead tip with transmit probe and the IPG end. (B) RF heating experiment setup showing

phantom and implant configurations in 0.55 T Scanner. ASTM phantom filled with tissue mimicking gel and setup for an example lead configuration for RF heating experiment also shown.

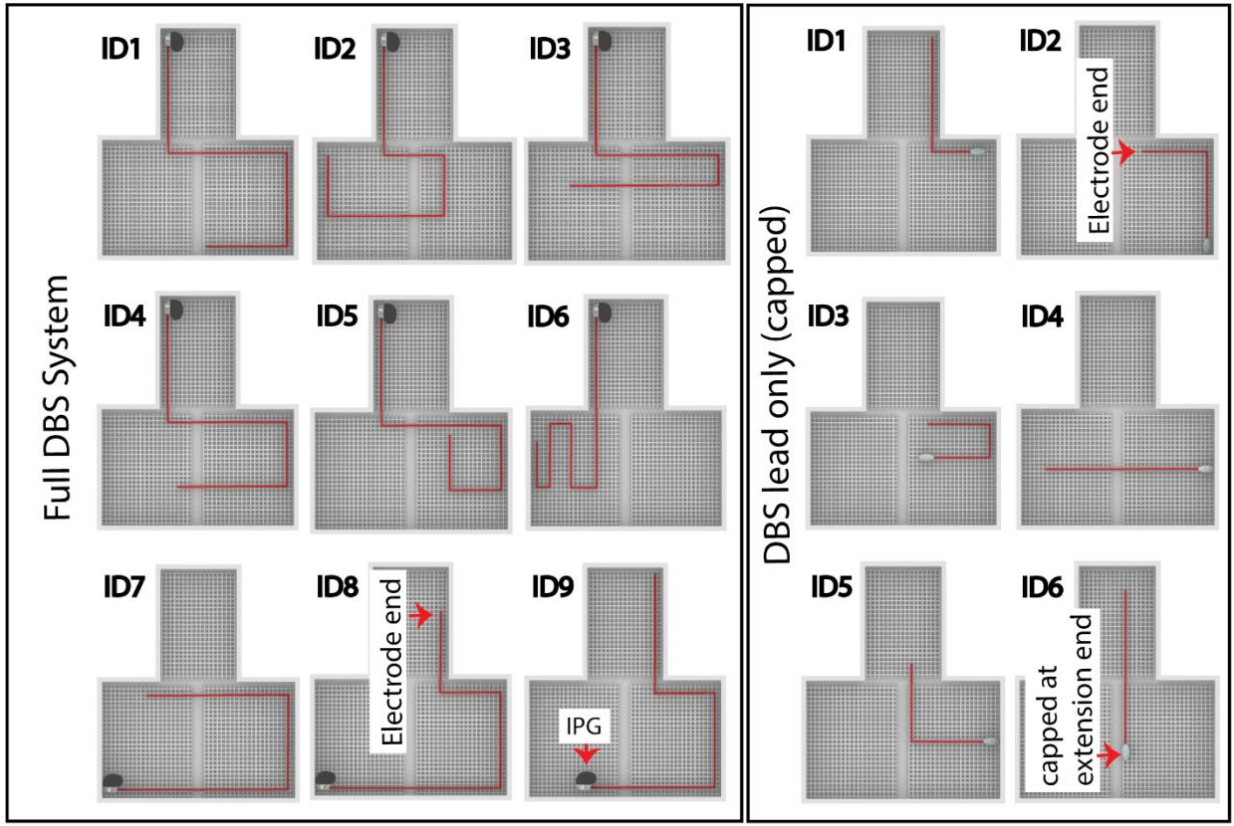

Figure 2: Trajectories of a full DBS device as well as lead only (capped), used for evaluation of temperature increase during MR imaging at 0.55 T and 1.5 T.

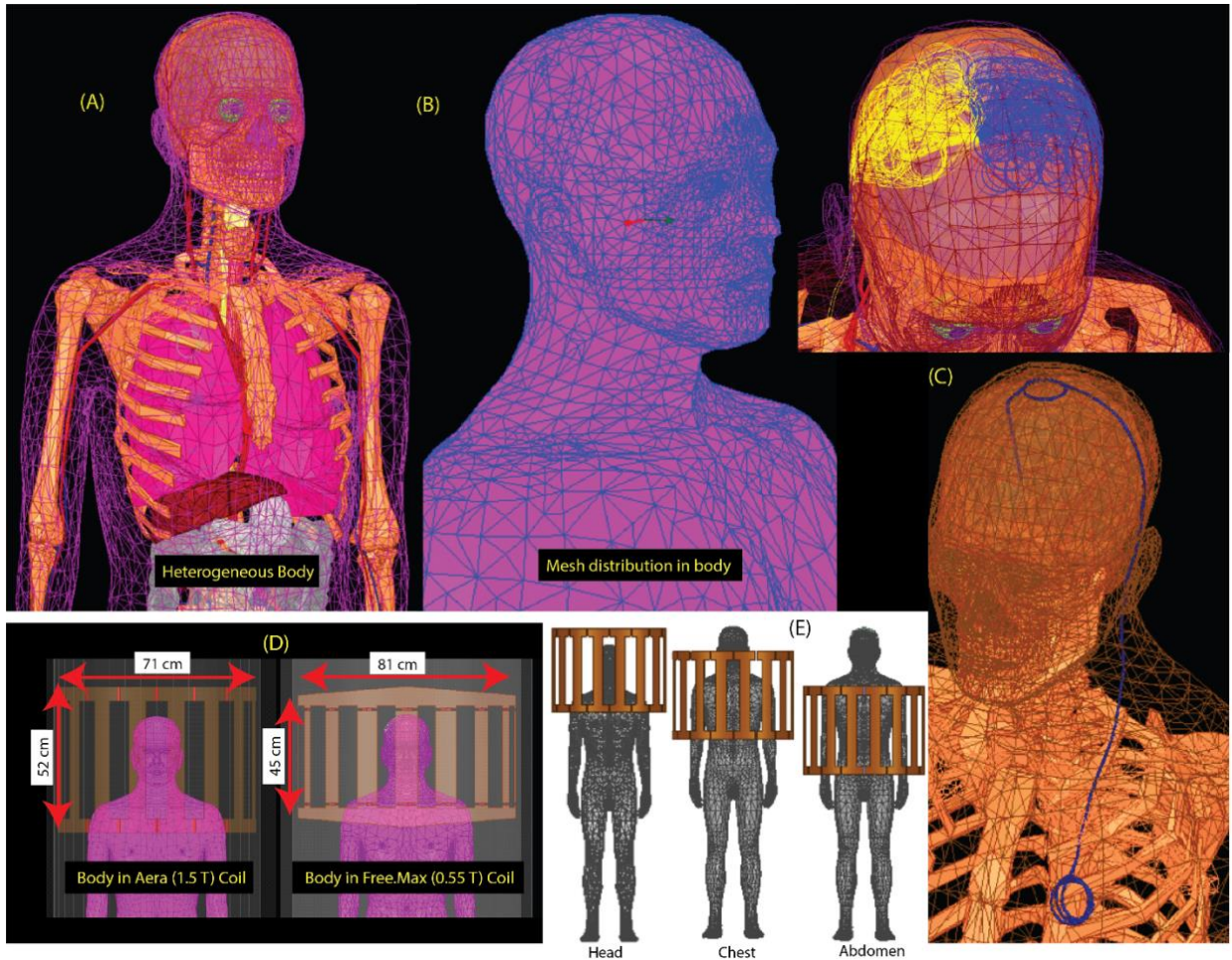

Figure 3: (A & B) Simulation setup showing heterogenous body model with different tissue types, and mesh distribution in the body. (C) The DBS trajectories overlaid on the skull of the human model, representing ipsilateral (yellow) and contralateral (blue) DBS implantation. A typical full DBS trajectory configuration also shown. (D) Body model positioned at head imaging landmark inside the 1.5 T Aera coil and 0.55 T Free.Max coil. (E) Different position of the body inside the RF coil to represent the RF exposure during MRI at head, chest and abdomen imaging landmarks.

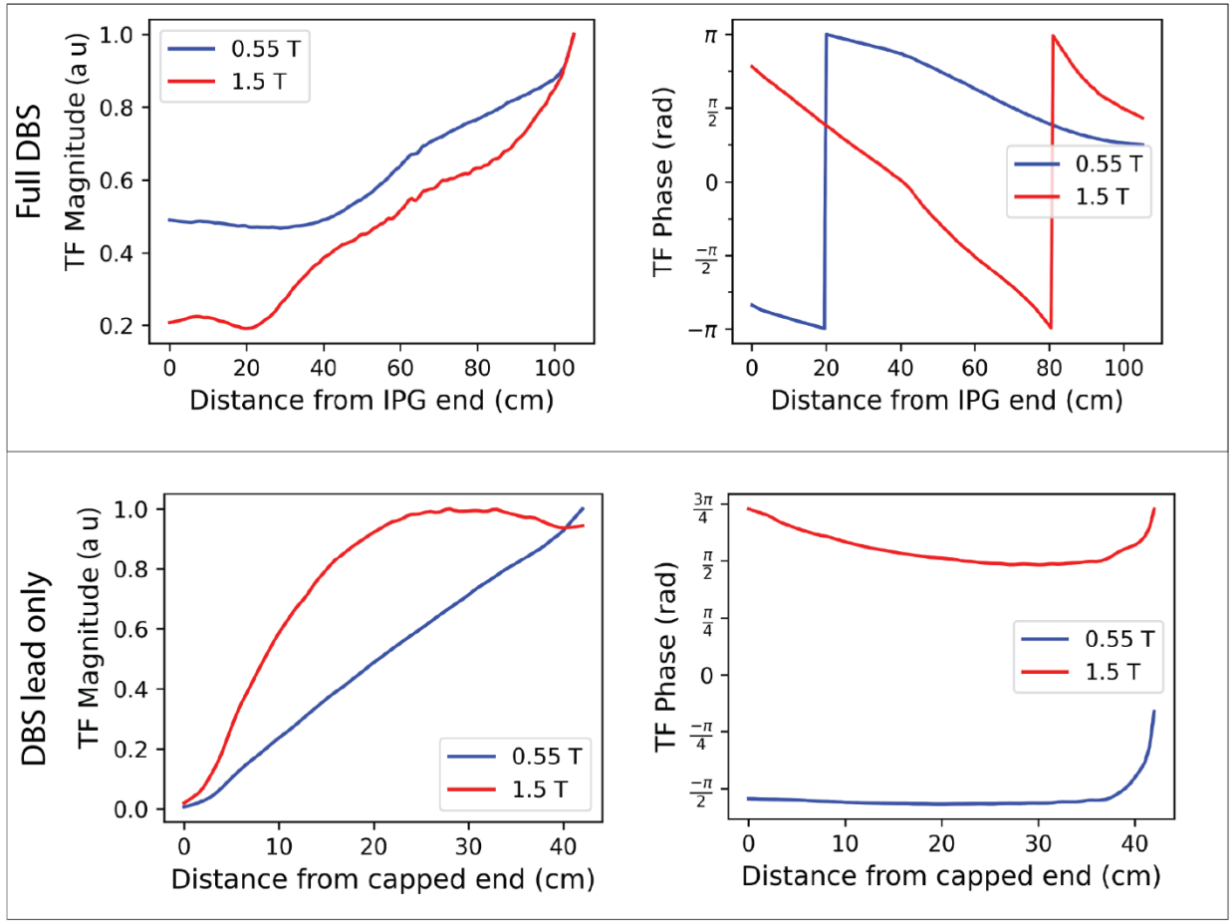

Figure 4: Magnitude and phase of transfer function of full DBS device as well as DBS lead only cases, measured at 23.6 MHz and 63.6 MHz, representing 0.55 T and 1.5 T MRI scanners respectively. The TF magnitude for each case has been normalized to its maximum value and is shown in arbitrary units (a u).

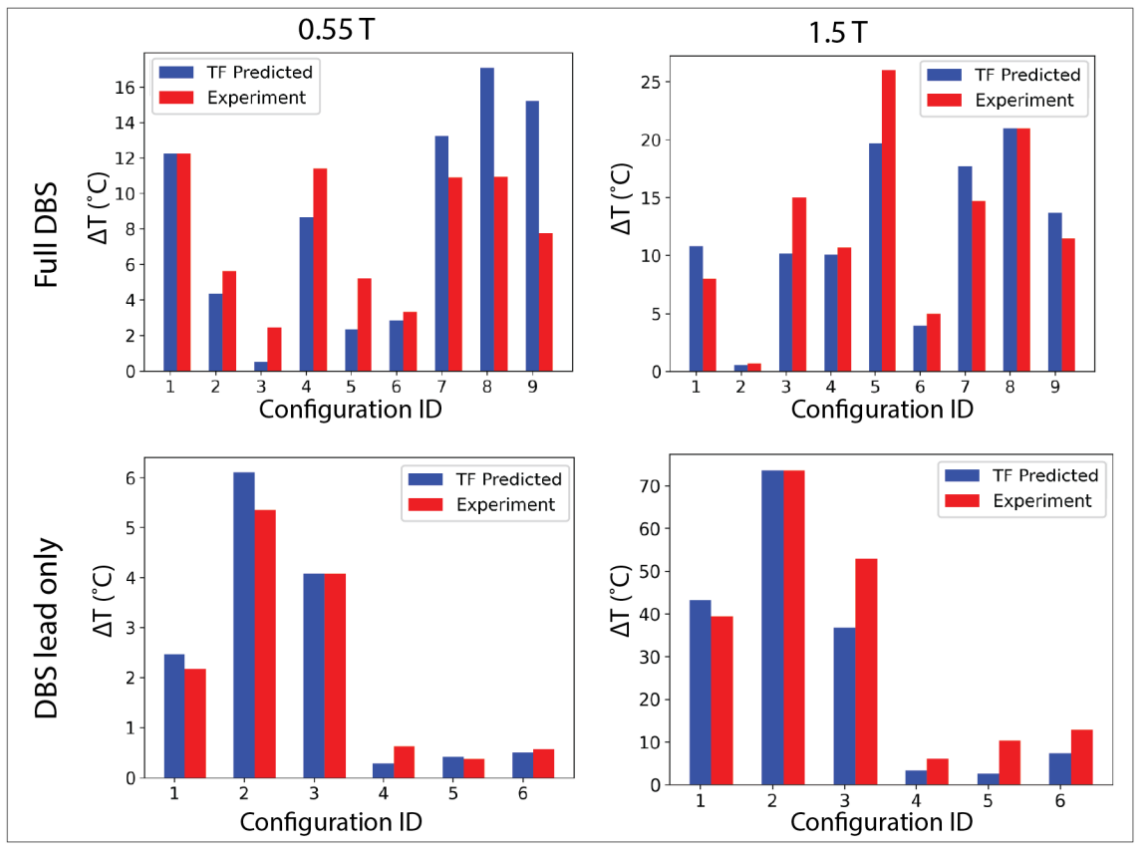

Figure 5: Bar plots showing comparison of the measured and AIMD model predicted temperature increase at the DBS contact during MR imaging on the 0.55 T and 1.5 T scanners. The plots are shown for both the full DBS device as well as the DBS lead-only system.

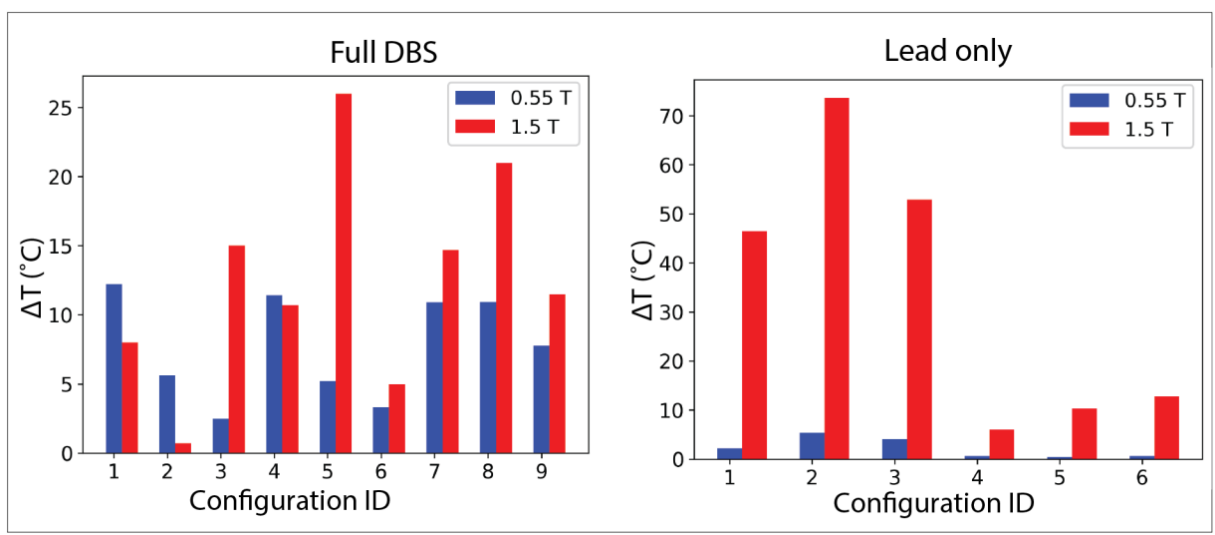

Figure 6: Bar plots showing comparison of temperature increase for full DBS system as well as capped lead only, measured from phantom experiment during MR imaging on the 0.55 T and 1.5 T scanners.

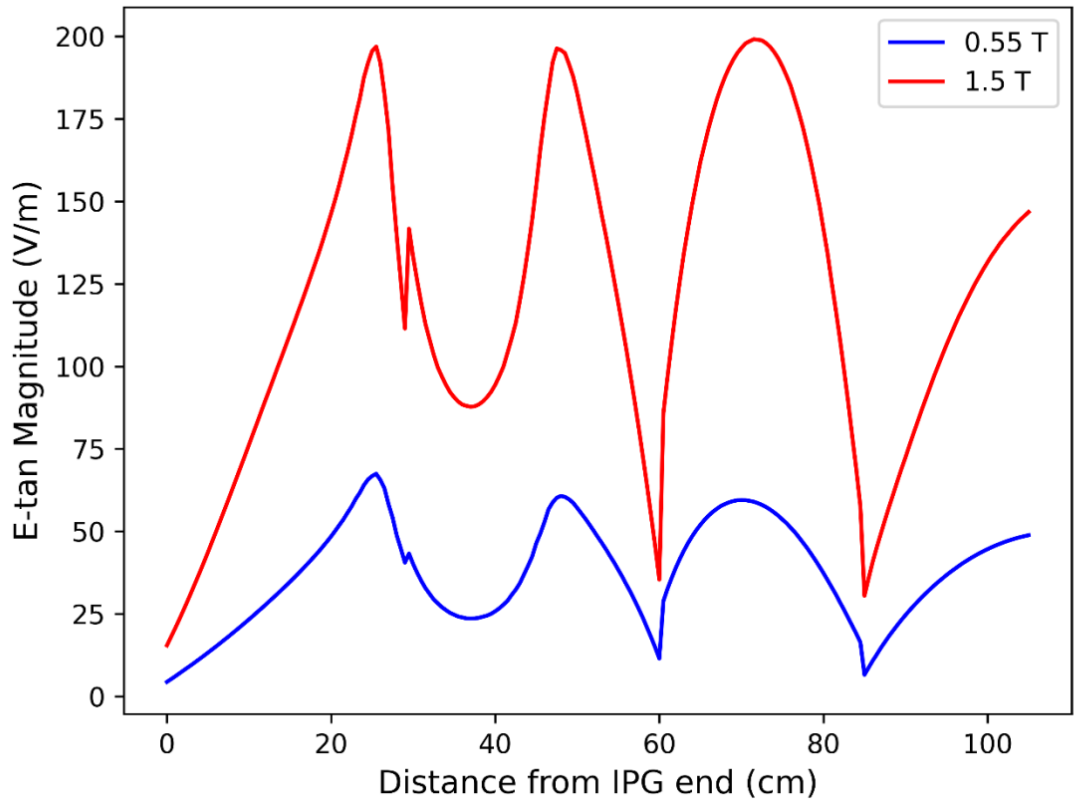

Figure 7: Plot for the magnitude of tangential component of E-field along the configuration ID 1 (see Figure 2) of full DBS device obtained from simulation when coil power for both 1.5 T and 0.55 T coils were adjusted to produce $\mathbf{B_1^+}$rms of 4.5 µT.

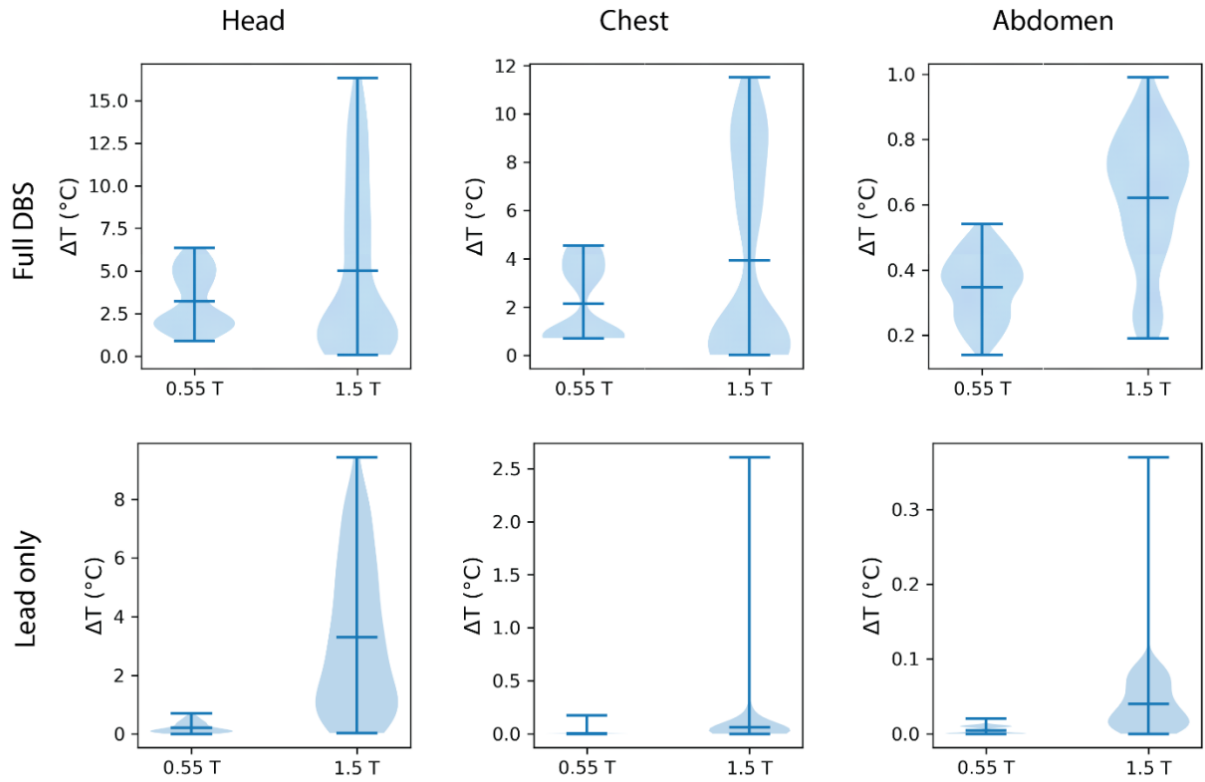

Figure 8: Violin plots showing the distribution of predicted temperature increase for the full DBS system as well as the lead-only case during RF exposure at 0.55 T and 1.5 T for head, chest and abdomen imaging landmarks.

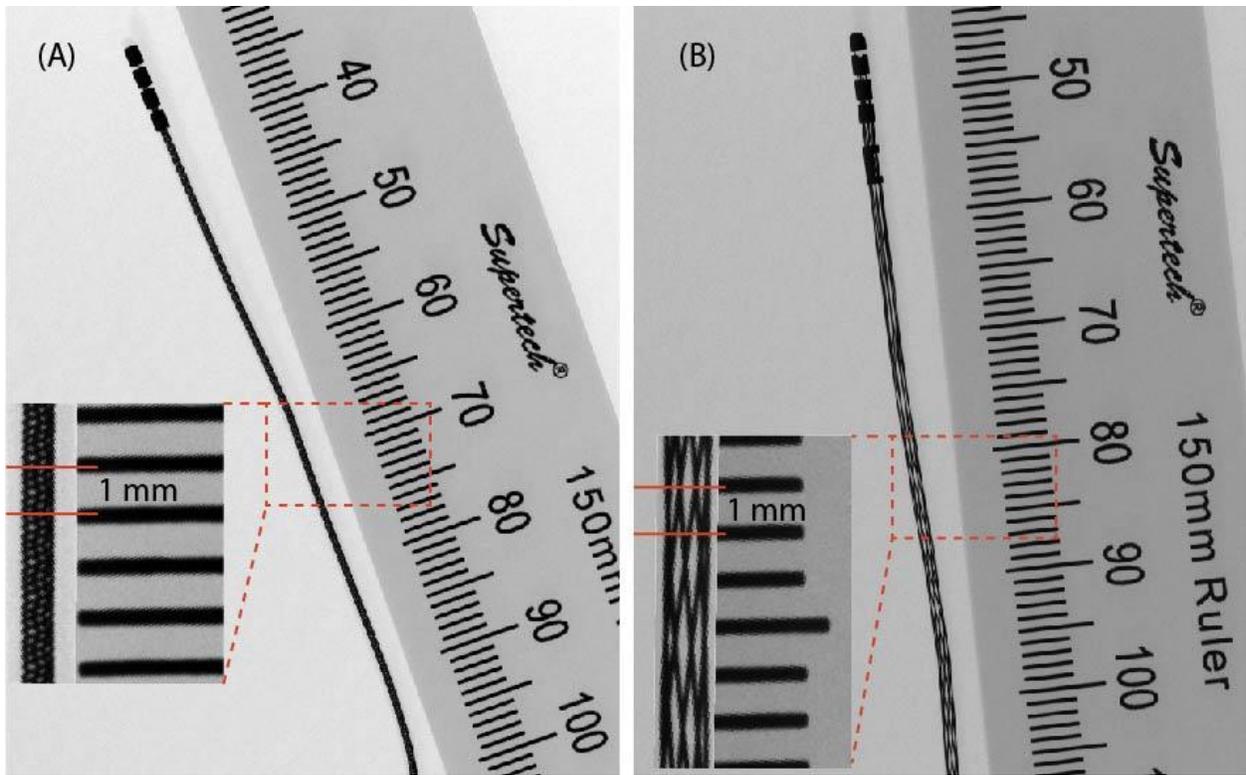
Figure 9: Mammograms of DBS leads from two different manufacturers, showing differences in the internal geometries, especially highlighting the difference in the pitch of the helix wound from the conductors connected to each of the electrode contact.

**Supplementary File:**

Table S1: Dielectric properties of different tissue classes from the heterogenous human model used for the in vivo prediction of RF heating at frequencies of 23.6 MHz and 63.6 MHz.

| Tissue | Conductivity (S/m) | | Relative Permittivity | |
| --- | --- | --- | --- | --- |
| | 23.6 MHz | 63.6 MHz | 23.6 MHz | 63.6 MHz |
| Blood | 1.15 | 1.2 | 141 | 87 |
| Skin | 0.32 | 0.36 | 91 | 64 |
| Bone (cancellous) | 0.14 | 0.16 | 45 | 31 |
| Bone (cortical) | 0.05 | 0.06 | 23.4 | 16.7 |
| Brain | 0.26 | 0.35 | 223 | 142 |
| Cartilage | 0.41 | 0.45 | 99 | 63 |
| Fat | 0.03 | 0.04 | 9 | 7 |
| Muscle | 0.65 | 0.69 | 105 | 72 |
| Heart | 0.57 | 0.68 | 174 | 107 |
| Lungs | 0.25 | 0.29 | 64 | 37 |
| Liver | 0.37 | 0.45 | 131 | 81 |
| Stomach | 0.83 | 0.88 | 131 | 86 |
| Small intestine | 1.46 | 1.49 | 231 | 119 |
| Nerve | 0.26 | 0.31 | 90 | 55 |
| Kidney | 0.61 | 0.74 | 208 | 119 |
| Colon | 0.55 | 0.64 | 154 | 95 |

Table S2: Values of predicted in-vivo temperature increase for Full-DBS as well as lead-only cases showing 95% confidence intervals for imaging at different landmarks with different values of $B_1^+$rms. For the lead-only case, only head landmark values are shown as the chest and abdomen landmark produced negligible heating.

| $B_1^+$ rms (µT) | 95% confidence interval of temperature increase (°C) | | | | | | | |
|---|---|---|---|---|---|---|---|---|
| | Full DBS | | | | | | DBS lead only (capped) | |
| | 0.55 T | | | 1.5 T | | | 0.55 T | 1.5 T |
| | Head | Chest | Abdomen | Head | Chest | Abdomen | Head | Head |
| 0.5 | 0.01 - 0.08 | 0.01 - 0.06 | 0.00 - 0.01 | 0.00 - 0.19 | 0.00 - 0.13 | 0.00 - 0.01 | 0.00 - 0.01 | 0.00 - 0.01 |
| 1.0 | 0.05 - 0.30 | 0.04 - 0.22 | 0.01 - 0.03 | 0.01 - 0.74 | 0.01 - 0.53 | 0.01 - 0.04 | 0.00 - 0.03 | 0.01 - 0.41 |
| 1.5 | 0.12 - 0.68 | 0.08 - 0.50 | 0.02 - 0.06 | 0.03 - 1.67 | 0.03 - 1.20 | 0.02 - 0.10 | 0.00 - 0.07 | 0.03 - 0.91 |
| 2.0 | 0.22 - 1.21 | 0.15 - 0.88 | 0.03 - 0.10 | 0.05 - 2.94 | 0.05 - 2.13 | 0.04 - 0.18 | 0.00 - 0.12 | 0.05 - 1.62 |
| 2.5 | 0.34 - 1.89 | 0.23 - 1.38 | 0.05 - 0.16 | 0.08 - 4.64 | 0.08 - 3.33 | 0.07 - 0.28 | 0.00 - 0.19 | 0.08 - 2.54 |
| 3.0 | 0.49 - 2.72 | 0.33 - 1.98 | 0.08 - 0.23 | 0.12 - 6.68 | 0.12 - 4.79 | 0.10 - 0.40 | 0.00 - 0.27 | 0.11 - 3.65 |
| 3.5 | 0.67 - 3.71 | 0.45 - 2.70 | 0.1 - 0.31 | 0.16 - 9.09 | 0.16 - 6.52 | 0.13 - 0.55 | 0.01 - 0.37 | 0.15 - 4.97 |
| 4.0 | 0.87 - 4.84 | 0.58 - 3.52 | 0.14 - 0.41 | 0.21 - 11.87 | 0.21 - 8.52 | 0.18 - 0.72 | 0.01 - 0.48 | 0.20 - 6.50 |

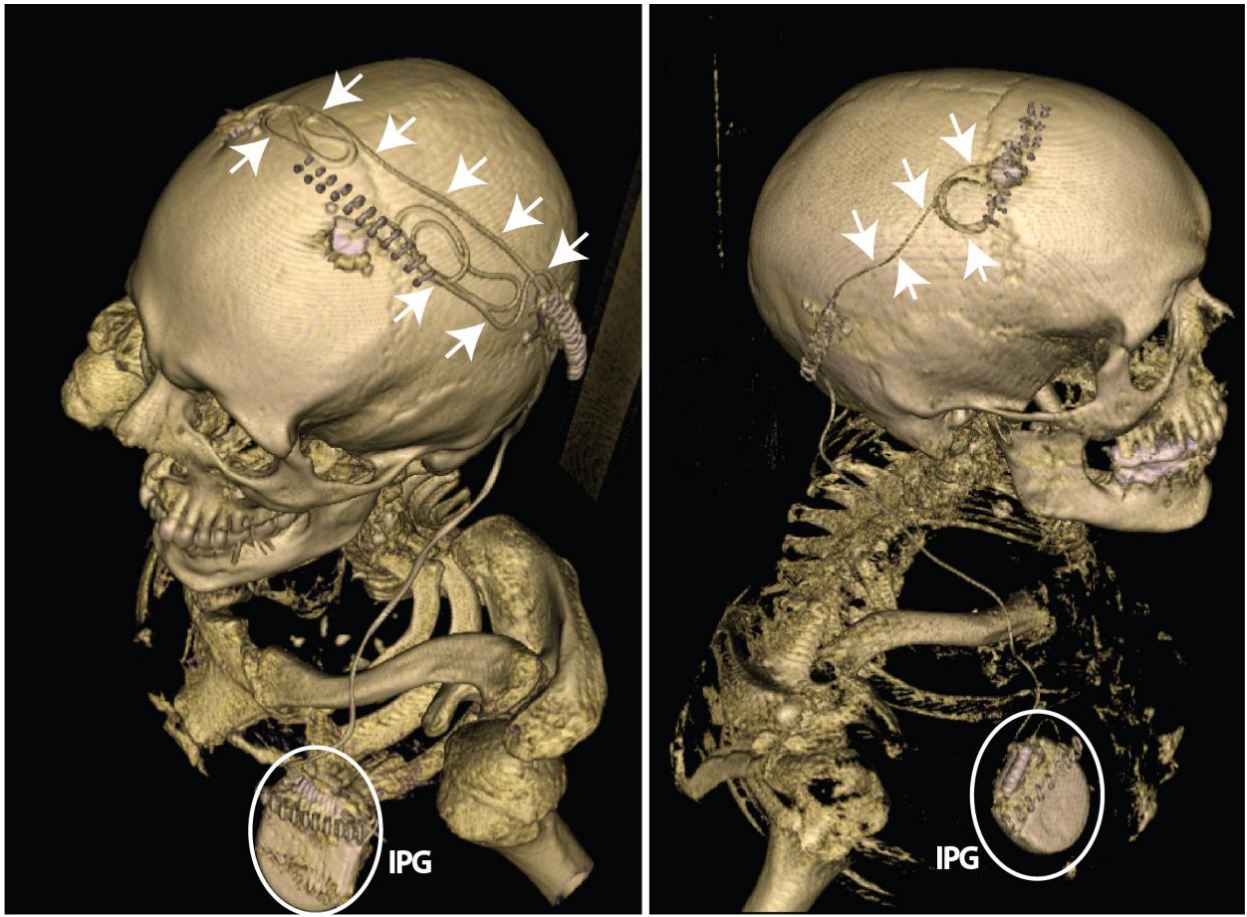

Figure S1: Volume rendered images of DBS patients with unilateral as well as bilateral DBS implants. The extracranial portion of the leads is shown by arrows, and the IPG is highlighted by circle mark.

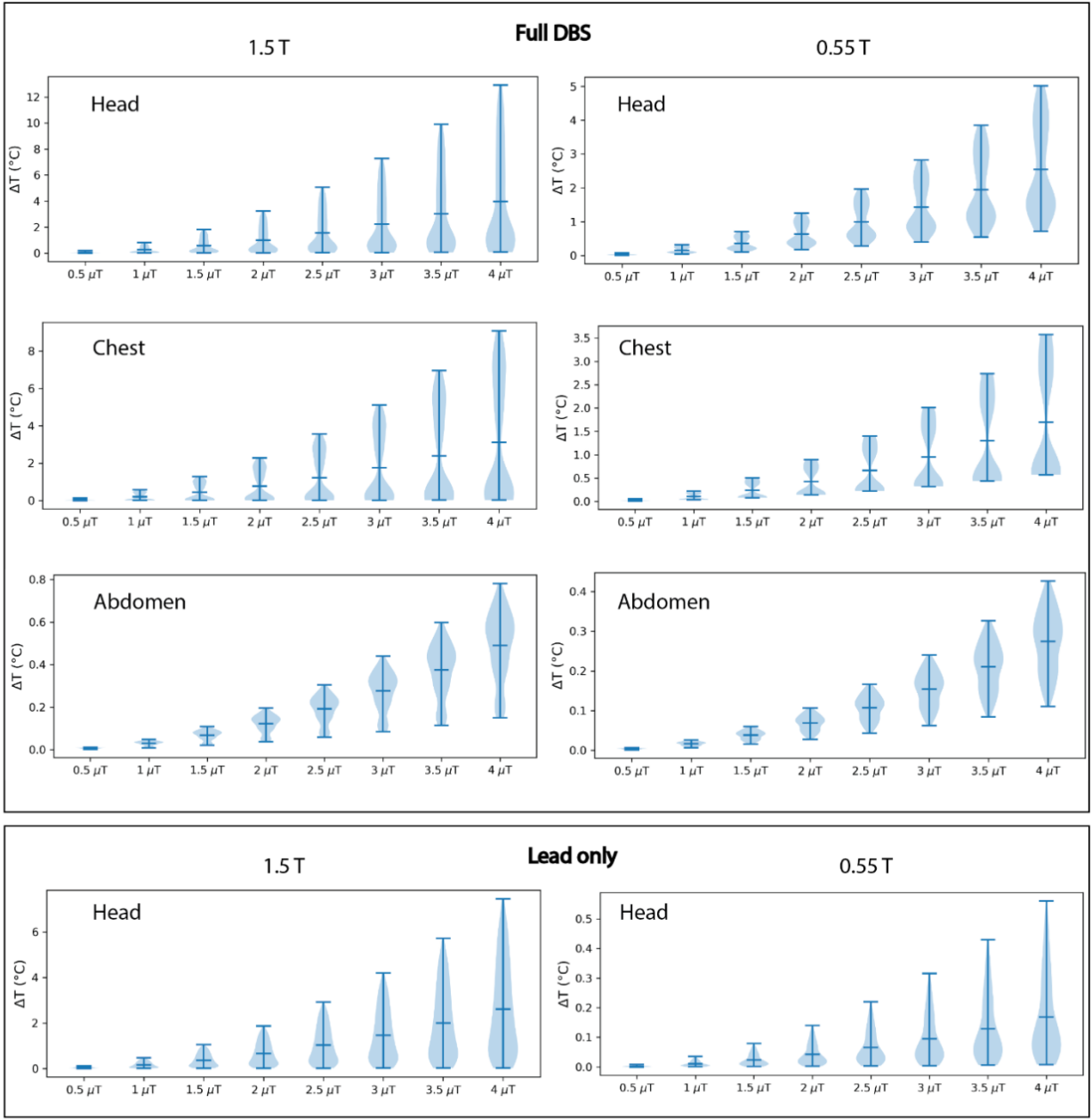

Figure S2: Violin plots showing predicted temperature increases for different values of $B_1^+$ rms for each of the imaging landmarks during MR imaging at 0.55 T and 1.5 T. For the lead-only case, only head imaging data are provided since the heating at the chest and abdomen landmarks was negligible.